\documentclass[12pt]{article}
\usepackage{latexsym,epsfig,epic,eepic}
\newcommand{\beq}{\begin{equation}}
\newcommand{\eeq}{\end{equation}}
\newcommand{\beqs}{\begin{eqnarray}}
\newcommand{\eeqs}{\end{eqnarray}}
\newcommand{\tr}{\mathrm{tr}}
\newcommand{\word}[2]{\left\{ \!\!\begin{array}{c} #1 \\ #2 \end{array}\!\!
\right\}}
\begin{document}
\begin{titlepage}
\vskip 2.5cm
\begin{center}
{\LARGE \bf Higher Charges in Dynamical}\\
\bigskip
{\LARGE \bf Spin Chains for SYM theory}\\
\vspace{2.71cm} {\large Abhishek Agarwal$^1$ and Gabriele
Ferretti$^2$} \vskip 0.7cm {\large \it $^1$ University of
Rochester\\ \smallskip Department of Physics and Astronomy,
Rochester NY 14627 USA} \\
{\large \it$^2$ Institute of Fundamental Physics \\
\smallskip
Chalmers University of Technology, 412 96 G\"oteborg, Sweden}
\vskip 0.3cm
\end{center}
\vspace{3.14cm}

\begin{abstract}
We construct, to the first two non-trivial orders, the next conserved
charge in the $su(2|3)$ sector of $\mathcal{N}=4$ Super Yang-Mills
theory. This represents a test of integrability in a sector where
the interactions change the number of sites of the chain. The
expression for the charge is completely determined by the algebra
and can be written in a diagrammatic form in terms of the
interactions already present in the Hamiltonian. It appears likely
that this diagrammatic expression remains valid in the full theory
and can be generalized to higher loops and higher charges thus
helping in establishing complete integrability for these
dynamical chains.
\end{abstract}

\end{titlepage}

\section{Introduction and Summary}
Recently, much effort has been devoted to the study of maximally
supersymmetric ($\mathcal{N}=4$) Super Yang--Mills theory (SYM) as
a first step toward uncovering some of the mysteries shrouding
strongly coupled gauge theories such as QCD. One very fruitful
line of investigation, pioneered by the work of Minahan and
Zarembo~\cite{Minahan:2002ve}, is to employ the fact that the
matrix of planar anomalous dimensions of local operators of this
gauge theory can be though of as the Hamiltonian of a quantum spin
chain. In~\cite{Minahan:2002ve} the authors considered the scalar
sector of the theory at first order in perturbation theory,
but the result was quickly generalized (still at one loop) to the
complete set of operators in~\cite{Beisert:2003yb}. The complete
one loop dilatation operator of $\mathcal{N} = 4$ SYM is a $psu(2,2|4)$
invariant spin chain, which turns out to be
integrable\footnote{Other aspects of integrability had
already appeared in the context of
QCD~\cite{Bukhvostov:1985rn,Lipatov:1993yb,Faddeev:1994zg}.
See~\cite{Braun:2003rp,Ferretti:2004ba,Belitsky:2004cz,
Belitsky:2004sc,Beisert:2004fv} for further work along these lines.}. 
Integrability
introduces an extremely powerful tool in the arsenal of gauge
theory computations, as it allows one to explicitly diagonalize
the matrix of anomalous dimensions
by the method of the Bethe ansatz. Although one does not
yet have a complete understanding of the dilatation
operator, by now the latter is known  to rather high orders in perturbation
theory in various closed sub-sectors of the gauge theory.

The particular sub-sector that will be analyzed in this paper
is the so-called $su(2|3)$ sector~\cite{Beisert:2003ys}, formed out of
a two-component fermion and three scalars, which is known to be closed to all
orders in perturbation theory. This sector inherits
a global $su(2|3)$($\in $ $psu(2,2|4)$) symmetry, and it has
proved to be a useful testing ground for ideas related
to higher loop integrability in the complete gauge theory. From
the work of Beisert~\cite{Beisert:2003ys}, the dilatation operator
is known to the third order in perturbation theory in this sector.

Work has also been done on a $su(2)$($\in $ $su(2|3)$) sub-sector,
where the corresponding three loop spin chain has been shown by
Serban and Staudacher~\cite{Serban:2004jf}  to be embedded in a
well known integrable long ranged spin chain, known as the
Inozemtsev spin chain. Extending this work, the all loop Bethe
equations and factorized scattering matrices have also been
proposed in~\cite{Beisert:2004hm} for the $su(2)$ sub-sector of
the gauge theory. The proposals for all loop Bethe ansatz and
factorized $S$ matrices have also been generalized to include the
$su(1|1), sl_2$, $su(1|2)$ and $psu(1,1|2)$
sub-sectors~\cite{Beisert:2005fw} (see also~\cite{Staudacher:2004tk}). 
Assuming that the proposed
$S$ matrices and Bethe equations correctly describe all these
sectors of the gauge theory, then, owing to the factorized nature
of the $S$ matrix, integrability would be obvious in these sectors
as well. Finally, also in~\cite{Beisert:2005fw} a remarkable ansatz was
proposed that generalizes all these results
to the full $psu(2,2|4)$ algebra. So far, no corresponding
factorized $S$ matrix is known that can incorporate the full
$psu(2,2|4)$ sector, hence, the Bethe equations proposed in
~\cite{Beisert:2005fw} assume the existence of an underlying
integrable spin chain. The full $psu(2,2|4)$ sector of the gauge
theory includes  $su(2|3)$ sector as well, within which the
dilatation operator is known to the third order in perturbation
theory. Since many of the novel features, such as the dynamical
nature of the spin chains that are present in the complete gauge
theory dilatation operator at higher loops are also present in the
$su(2|3) $ sub-sector, it is important to make further checks of
integrability for such closed sub-sectors.

Much of the intuition that goes into the construction of the all
loop Bethe ansatz for the various sectors of the gauge theory
comes from previous detailed studies of the $su(2)$ invariant
long ranged spin chains
~\cite{Haldane:1992sj,Bernard:1993uz,Bernard:1993va, Inozemtsev:2002vb}.
However, when one ventures into the $su(2|3)$ sector, one
encounters novel spin chains which do not seem to have been
studied in the past. Specifically, since the 'length' of the spin
chains (i.e. the number of SYM fields inside a single trace
operator) is no longer a good quantum number, one has to deal with
spin chains that do not preserve the number of sites. Such
dynamical chains are not included in any of the smaller sectors
mentioned above and it is important to check in what sense (if
any) the notion of integrability generalizes to these chains.

For a given quantum spin chain, integrability is in general quite
hard to establish. Ideally, one would like to establish
integrability of a spin chain Hamiltonian by relating it to a
transfer matrix satisfying a Yang-Baxter algebra, or by showing
that its scattering matrix is factorized. In all the cases for
which this is possible, one can show that the Hamiltonian is
member of a family of mutually commuting conserved charges. Hence,
given a spin chain that has a chance of being integrable, a test
of its integrability~\cite{Grabowski:1994rb} can be taken to be 
the existence of a
higher conserved charge whose rank (in the sense of the range of
interactions) is greater than that of the Hamiltonian 
(see also~\cite{Arutyunov:2003rg,Engquist:2004bx} for a 
$\sigma$-model perspective).

In the present context, the simplest way in which one can try
to generalize this notion of integrability to higher loops is for
the conserved charges $\mathcal{H}_s$ to be "deformed" by the
coupling constant $g = g_{YM}\sqrt{N_c}$, (the square root of
the 't Hooft coupling) as
$\mathcal{H}_s\to\mathcal{H}_s(g)$ in such a way that the mutual
commutators are still vanishing, at least when Taylor expanded
around the origin~\cite{Beisert:2003tq}.
This scenario has been referred to as
`perturbative integrability' in~\cite{Beisert:2003ys}. Precisely
in~\cite{Beisert:2003ys} the Hamiltonian  for the $su(2|3)$ sector of the
theory  was computed to three loops
and an argument for integrability, based on the degeneracy
of the parity pairs, was given. The one loop
$su(2|3)$ spin chain Hamiltonian is of the type that has been
studied in the condensed matter physics literature in the context
of the Hubbard model~\cite{Essler:1992uc}, and it can be regarded
as a supersymmetric generalization of the celebrated Heisenberg
Hamiltonian with nearest neighbor exchange interactions. The first
non-trivial conserved charge of this one-loop Hamiltonian (and
indeed all the higher ones) are known in a very explicit form.
Hence, a test of perturbative integrability in the $su(2|3)$ sector
consists in constructing a
$g$ dependent deformation of the next conserved
charge.

In the present work we perform this test and construct such
deformation to the first two non-trivial orders in $g$. This new conserved
charge is completely determined and can be expressed in a very simple
graphic form as a quadratic combination of terms in the Hamiltonian that
should easily generalize to higher loops, higher charges and, most importantly,
different integrable systems, including the full $psu(2,2|4)$ algebra.
In the light of
discussions such as~\cite{Grabowski:1994rb} this is a very strong check that
integrability does indeed survive in the dynamical chain.

\section{Test of Perturbative Integrability:}
"Perturbative Integrability" is the statement that for small values
of the 't Hooft coupling $g = g_{YM}\sqrt{N_c}$
there exists an infinite\footnote{ Or, for a chain of finite
length $L$, of order $L$} set of mutually commuting conserved
charges $\mathcal{H}_s(g)$ (we will take $s\geq 2$) analytic at $g=0$.
Performing a Taylor expansion on the charges:
\beq
     \mathcal{H}_s(g) = \sum_{k=0}^\infty  \mathcal{H}_{s,k} g^k
\eeq
we can expand the commutator of two charges to obtain an infinite set of
relations
\beq
    \sum_{k=0}^l {[ \mathcal{H}_{r,k},\mathcal{H}_{s,l-k}]}=0.
    \label{pertcom}
\eeq The first relation obtained from (\ref{pertcom}) for $l=0$ is the usual
relation discussed in the context of integrable spin chains:
\beq
{[ \mathcal{H}_{r,0},\mathcal{H}_{s,0}]}=0.
\eeq
The lowest charge $\mathcal{H}_2(g)$ corresponds to the dilatation
operator, to be thought of as the "Hamiltonian" of the spin
system. More specifically, adopting the standard notation,
we write the total scaling dimension $\mathcal{D}$ of an operator as
\beq
    \mathcal{D} = \mathcal{D}_0 + g^2 \mathcal{H}_2(g),
\eeq where $\mathcal{D}_0$ is the classical (mass) dimension and
an overall factor of $g^2$ has been extracted form the charge.
Thus, with this conventions, $\mathcal{H}_2(0) =
\mathcal{H}_{2,0}$ corresponds to the one loop matrix of anomalous
dimensions. Although the Taylor expansion of $\mathcal{H}_2(g)$
may contain odd powers of $g$, it is possible to check that the
eigenvalues of $\mathcal{D}$  always contain even powers of $g$.

As mentioned in the introduction, it is crucial to investigate
sectors where the length of the chain is allowed to vary to test
the validity of (\ref{pertcom}) in the context of $\mathcal{N}=4$
SYM theory. The simplest such sector, closed to all loops, is the
so-called $su(2|3)$ sector, consisting of three bosons $\Phi^a$
($a=1,2,3$) and two fermions $\Psi^\alpha$ ($\alpha=1,2$). For
such sector, the Hamiltonian $\mathcal{H}_2(g)$ is known to order
$\mathcal{O}(g^4)$ from the work of Beisert~\cite{Beisert:2003ys}
and a natural question is to try to construct
the next charge $\mathcal{H}_3(g)$ to some order.
To zeroth order the charge $\mathcal{H}_{3,0} = \mathcal{H}_3(0)$
is already known to exist because the system is nothing but the
usual integrable
system with Hamiltonian described by (graded) permutations of
nearest neighbors and the higher charges at $g=0$ are easily
extracted from the transfer matrix. In this note we present the explicit
expression for the first non-trivial term in the deformation
$\mathcal{H}_{3,1}$. (See Appendix for $\mathcal{H}_{3,2}$.) 
Such term already mixes chains of different lengths
and thus represent a novel test of integrability.

\begin{figure}[hbp]
  \begin{center}
\setlength{\unitlength}{0.00066667in}
\begingroup\makeatletter\ifx\SetFigFont\undefined%
\gdef\SetFigFont#1#2#3#4#5{%
  \reset@font\fontsize{#1}{#2pt}%
  \fontfamily{#3}\fontseries{#4}\fontshape{#5}%
  \selectfont}%
\fi\endgroup%
{\renewcommand{\dashlinestretch}{30}
\begin{picture}(3024,1689)(0,-10)
\put(1362,837){\ellipse{600}{300}}
\path(12,1662)(12,12)
\path(237,1662)(237,12)
\path(462,1662)(462,12)
\path(687,1662)(687,12)
\path(912,1662)(912,12)
\path(1887,1662)(1887,12)
\path(2112,1662)(2112,12)
\path(2337,1662)(2337,12)
\path(2562,1662)(2562,12)
\path(2787,1662)(2787,12)
\path(3012,1662)(3012,12)
\path(1137,912)(1137,1662)
\path(1362,987)(1362,1662)
\path(1587,912)(1587,1662)
\path(1212,687)(1212,12)
\path(1512,687)(1512,12)
\end{picture}
}
 \end{center}
  \caption{{\it Graphical representation of an
interaction that does not conserve
the length of the chain. Each line ends on one site of a chain.
The interaction, denoted by a blob, is supposed to be moved (summed) over
the chain. For $\mathcal{N}=4$ SYM one considers only periodic chains,
i.e. identifies the ends of the diagram.}}
  \label{exa}
\end{figure}
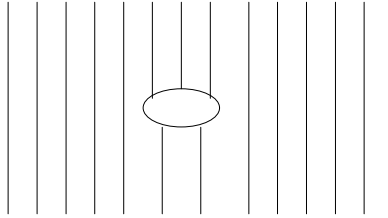

In presenting the explicit expressions for these charges, we will use
Beisert's notation~\cite{Beisert:2003ys}
to denote the local interactions that, when summed over
the chain, give rise to the charges.
Briefly, an arbitrary $su(2|3)$ spin chain can be written as
$\tr(W^{A_1}\dots W^{A_n})$, where $W^A = \Phi^a~\mathrm{or}~\Psi^\alpha$,
that is, we use capital letters to collectively
describe all five spin states, while lower case Latin letters and
Greek letters describe the three Bosonic and the
two fermionic values of the spins respectively. A generic local interaction
can be written as
\beq
    \word{A_1\dots A_p}{B_1\dots B_q}
\eeq
and can be thought of as a ``machine'' that runs along the chain and
whenever it finds the combination of indices $A_1\dots A_p$ it replaces it
with the combination $B_1\dots B_q$. In a ``dynamical'' chain the number
of sites need not be conserved, i.e. $p\not=q$ (Fig.~\ref{exa}).

In the following,
we will only need to know the expression for $\mathcal{H}_{2,0}$ and
$\mathcal{H}_{2,1}$. (The latter, together with $\mathcal{H}_{2,2}$,
is used to determine the two loop anomalous dimension.) We will also
need the expressions for the supercharge $\mathcal{Q}_\alpha^a(g)$ to the
same order. (The other supercharge, usually denoted by
$\mathcal{S}^\alpha_a(g)$ will not give further constraints
on $\mathcal{H}_3(g)$.) The diagrammatic expression for $\mathcal{H}_2(g)$
and $\mathcal{H}_3(g)$ are given in Fig.~\ref{charge2} and Fig.~\ref{charge3}
respectively
\begin{figure}[hbp]
  \begin{center}
\setlength{\unitlength}{0.00066667in}
\begingroup\makeatletter\ifx\SetFigFont\undefined%
\gdef\SetFigFont#1#2#3#4#5{%
  \reset@font\fontsize{#1}{#2pt}%
  \fontfamily{#3}\fontseries{#4}\fontshape{#5}%
  \selectfont}%
\fi\endgroup%
{\renewcommand{\dashlinestretch}{30}
\begin{picture}(8292,1389)(0,-10)
\put(3300,687){\ellipse{600}{300}}
\put(2100,687){\ellipse{600}{300}}
\put(825,687){\ellipse{600}{300}}
\put(4875,687){\ellipse{600}{300}}
\put(5925,687){\ellipse{600}{300}}
\put(6975,687){\ellipse{600}{300}}
\path(1950,1362)(1950,837)
\path(1950,1362)(1950,837)
\path(2100,1362)(2100,837)
\path(2100,1362)(2100,837)
\path(2250,1362)(2250,837)
\path(2250,1362)(2250,837)
\path(2175,537)(2175,12)
\path(2175,537)(2175,12)
\path(2025,537)(2025,12)
\path(2025,537)(2025,12)
\path(3150,537)(3150,12)
\path(3150,537)(3150,12)
\path(3300,537)(3300,12)
\path(3300,537)(3300,12)
\path(3450,537)(3450,12)
\path(3450,537)(3450,12)
\path(3225,1362)(3225,837)
\path(3225,1362)(3225,837)
\path(3375,1362)(3375,837)
\path(3375,1362)(3375,837)
\path(750,537)(750,12)
\path(750,537)(750,12)
\path(900,537)(900,12)
\path(900,537)(900,12)
\path(900,1362)(900,837)
\path(900,1362)(900,837)
\path(750,1362)(750,837)
\path(750,1362)(750,837)
\path(4650,762)(4650,1287)
\path(4650,762)(4650,1287)
\path(4800,1362)(4800,837)
\path(4800,1362)(4800,837)
\path(4950,1362)(4950,837)
\path(4950,1362)(4950,837)
\path(5100,1287)(5100,762)
\path(5100,1287)(5100,762)
\path(4800,537)(4800,12)
\path(4800,537)(4800,12)
\path(4950,537)(4950,12)
\path(4950,537)(4950,12)
\path(5775,1362)(5775,837)
\path(5775,1362)(5775,837)
\path(5925,1362)(5925,837)
\path(5925,1362)(5925,837)
\path(5775,537)(5775,12)
\path(5775,537)(5775,12)
\path(5925,537)(5925,12)
\path(5925,537)(5925,12)
\path(6075,537)(6075,12)
\path(6075,537)(6075,12)
\path(6075,1362)(6075,837)
\path(6075,1362)(6075,837)
\path(6900,1362)(6900,837)
\path(6900,1362)(6900,837)
\path(7050,1362)(7050,837)
\path(7050,1362)(7050,837)
\path(6750,612)(6750,87)
\path(6750,612)(6750,87)
\path(6900,537)(6900,12)
\path(6900,537)(6900,12)
\path(7050,537)(7050,12)
\path(7050,537)(7050,12)
\path(7200,612)(7200,87)
\path(7200,612)(7200,87)
\put(0, 700){\makebox(0,0){$\mathcal{H}_2 = $}}
\put(1475,680){\makebox(0,0){$+ g \;\big($}}
\put(4100,680){\makebox(0,0){$\big)+ g^2\;\big($}}
\put(2690,680){\makebox(0,0){$+$}}
\put(5400,680){\makebox(0,0){$+$}}
\put(6450,680){\makebox(0,0){$+$}}
\put(7730,680){\makebox(0,0){$\big)+\dots$}}
\end{picture}
}
 \end{center}
  \caption{{\it First conserved charge $\mathcal{H}_2(g)$ (Hamiltonian).
The fourth and sixth blob
are identically zero in the $su(2|3)$ sector.}}
  \label{charge2}
\end{figure}
\vfill\eject
Thus, we take as our Hamiltonian, to zeroth and first
order~\cite{Beisert:2003ys}:
\beqs
     \mathcal{H}_{2,0} &=& \word{AB}{AB}-(-1)^{(A)(B)}\word{AB}{BA}
     \nonumber \\
     &=& \word{ab}{ab} + \word{\alpha b}{\alpha b} +
    \word{a\beta}{a\beta} + \word{\alpha\beta}{\alpha\beta} - \nonumber \\
    && \word{ab}{ba} - \word{\alpha b}{b\alpha} -
    \word{a\beta}{\beta a} + \word{\alpha\beta}{\beta\alpha},
   \nonumber \\
     \mathcal{H}_{2,1}&=& \epsilon_{abc}\epsilon_{\alpha\beta}
     \word{abc}{\alpha\beta}+ \epsilon_{abc}\epsilon_{\alpha\beta}
     \word{\alpha\beta}{abc}\label{q2}.
\eeqs
We used the freedom allowed by a scaling of the coupling constant and a phase
rotation to fix all coefficients in (\ref{q2}) without loss of generality.
(In particular, we dropped a factor of $-1/\sqrt{2}$ in $\mathcal{H}_{2,1}$
that is required to get the right values for the anomalous dimensions 
but is irrelevant for our purpose.)
The symbol $(A)$ (and more generally
$(A_1\dots A_n)$) at the exponent is equal to zero if the quantity within
parenthesis is bosonic and equal to one if it is fermionic,
thus enforcing the grading of the permutation.

In the same spirit, we define the supercharges
\beq
      \mathcal{Q}^a_\alpha(g) = \sum_{k=0}^\infty 
      {\mathcal{Q}_k}^a_\alpha\, g^k,
\eeq
where
\beq
     {\mathcal{Q}_0}^a_\alpha = \word{a}{\alpha} \quad\mathrm{and}\quad
     {\mathcal{Q}_1}^a_\alpha =
      \epsilon_{abc}\epsilon_{\alpha\beta}\word{\beta}{bc}
\eeq
\begin{figure}[h!bp]
  \begin{center}
\setlength{\unitlength}{0.00066667in}
\begingroup\makeatletter\ifx\SetFigFont\undefined%
\gdef\SetFigFont#1#2#3#4#5{%
  \reset@font\fontsize{#1}{#2pt}%
  \fontfamily{#3}\fontseries{#4}\fontshape{#5}%
  \selectfont}%
\fi\endgroup%
{\renewcommand{\dashlinestretch}{30}
\begin{picture}(8067,1389)(0,-10)
\texture{88555555 55000000 555555 55000000 555555 55000000 555555 55000000
    555555 55000000 555555 55000000 555555 55000000 555555 55000000
    555555 55000000 555555 55000000 555555 55000000 555555 55000000
    555555 55000000 555555 55000000 555555 55000000 555555 55000000 }
\put(3300,687){\shade\ellipse{600}{300}}
\put(3300,687){\ellipse{600}{300}}
\put(4875,687){\shade\ellipse{600}{300}}
\put(4875,687){\ellipse{600}{300}}
\put(5925,687){\shade\ellipse{600}{300}}
\put(5925,687){\ellipse{600}{300}}
\put(6975,687){\shade\ellipse{600}{300}}
\put(6975,687){\ellipse{600}{300}}
\put(823,691){\shade\ellipse{600}{300}}
\put(823,691){\ellipse{600}{300}}
\put(2084,703){\shade\ellipse{600}{300}}
\put(2084,703){\ellipse{600}{300}}
\path(675,1362)(675,837)
\path(675,1362)(675,837)
\path(975,1362)(975,837)
\path(975,1362)(975,837)
\path(675,537)(675,12)
\path(675,537)(675,12)
\path(975,537)(975,12)
\path(975,537)(975,12)
\path(825,1362)(825,837)
\path(825,1362)(825,837)
\path(825,537)(825,12)
\path(825,537)(825,12)
\path(1950,537)(1950,12)
\path(1950,537)(1950,12)
\path(2100,537)(2100,12)
\path(2100,537)(2100,12)
\path(1875,1287)(1875,762)
\path(1875,1287)(1875,762)
\path(2325,1287)(2325,762)
\path(2325,1287)(2325,762)
\path(2025,1362)(2025,837)
\path(2025,1362)(2025,837)
\path(2175,1362)(2175,837)
\path(2175,1362)(2175,837)
\path(3075,612)(3075,87)
\path(3075,612)(3075,87)
\path(3525,612)(3525,87)
\path(3525,612)(3525,87)
\path(3225,537)(3225,12)
\path(3225,537)(3225,12)
\path(3375,537)(3375,12)
\path(3375,537)(3375,12)
\path(3150,1362)(3150,837)
\path(3150,1362)(3150,837)
\path(3450,1362)(3450,837)
\path(3450,1362)(3450,837)
\path(3300,1362)(3300,837)
\path(3300,1362)(3300,837)
\path(4725,537)(4725,12)
\path(4725,537)(4725,12)
\path(5025,537)(5025,12)
\path(5025,537)(5025,12)
\path(4875,537)(4875,12)
\path(4875,537)(4875,12)
\path(4575,687)(4575,1212)
\path(4575,687)(4575,1212)
\path(4725,1287)(4725,762)
\path(4725,1287)(4725,762)
\path(5175,1212)(5175,687)
\path(5175,1212)(5175,687)
\path(5025,1287)(5025,762)
\path(5025,1287)(5025,762)
\path(4875,1362)(4875,837)
\path(4875,1362)(4875,837)
\path(5700,1287)(5700,762)
\path(5700,1287)(5700,762)
\path(5850,1362)(5850,837)
\path(5850,1362)(5850,837)
\path(6150,1287)(6150,762)
\path(6150,1287)(6150,762)
\path(6000,1362)(6000,837)
\path(6000,1362)(6000,837)
\path(5700,612)(5700,87)
\path(5700,612)(5700,87)
\path(5850,537)(5850,12)
\path(5850,537)(5850,12)
\path(6150,612)(6150,87)
\path(6150,612)(6150,87)
\path(6000,537)(6000,12)
\path(6000,537)(6000,12)
\path(6675,687)(6675,162)
\path(6675,687)(6675,162)
\path(6825,612)(6825,87)
\path(6825,612)(6825,87)
\path(7275,687)(7275,162)
\path(7275,687)(7275,162)
\path(7125,612)(7125,87)
\path(7125,612)(7125,87)
\path(6975,537)(6975,12)
\path(6975,537)(6975,12)
\path(6825,1362)(6825,837)
\path(6825,1362)(6825,837)
\path(7125,1362)(7125,837)
\path(7125,1362)(7125,837)
\path(6975,1362)(6975,837)
\path(6975,1362)(6975,837)
\path(2250,537)(2250,12)
\path(2250,537)(2250,12)
\put(0, 700){\makebox(0,0){$\mathcal{H}_3 = $}}
\put(1475,680){\makebox(0,0){$+ g \;\big($}}
\put(4100,680){\makebox(0,0){$\big)+ g^2\;\big($}}
\put(2690,680){\makebox(0,0){$+$}}
\put(5400,680){\makebox(0,0){$+$}}
\put(6450,680){\makebox(0,0){$+$}}
\put(7730,680){\makebox(0,0){$\big)+\dots$}}
\end{picture}
}
 \end{center}
  \caption{{\it Second conserved charge $\mathcal{H}_3(g)$.
The fourth and sixth blob
are identically zero in the $su(2|3)$ sector.}}
  \label{charge3}
\end{figure}

\begin{figure}[hbp]
  \begin{center}
\setlength{\unitlength}{0.00066667in}
\begingroup\makeatletter\ifx\SetFigFont\undefined%
\gdef\SetFigFont#1#2#3#4#5{%
  \reset@font\fontsize{#1}{#2pt}%
  \fontfamily{#3}\fontseries{#4}\fontshape{#5}%
  \selectfont}%
\fi\endgroup%
{\renewcommand{\dashlinestretch}{30}
\begin{picture}(4801,1689)(0,-10)
\put(2400,537){\ellipse{600}{300}}
\put(2700,1212){\ellipse{600}{300}}
\path(2250,387)(2250,12)
\drawline(2250,687)(2250,687)
\path(2250,687)(2250,1662)
\path(2475,387)(2475,12)
\path(2625,1062)(2475,687)
\path(2625,1362)(2625,1662)
\path(2850,1062)(2850,12)
\path(2850,1362)(2850,1662)
\put(3975,1137){\ellipse{600}{300}}
\put(4275,462){\ellipse{600}{300}}
\path(3825,1287)(3825,1662)
\drawline(3825,987)(3825,987)
\path(3825,987)(3825,12)
\path(4050,1287)(4050,1662)
\path(4200,612)(4050,987)
\path(4200,312)(4200,12)
\path(4425,612)(4425,1662)
\path(4425,312)(4425,12)
\texture{88555555 55000000 555555 55000000 555555 55000000 555555 55000000
    555555 55000000 555555 55000000 555555 55000000 555555 55000000
    555555 55000000 555555 55000000 555555 55000000 555555 55000000
    555555 55000000 555555 55000000 555555 55000000 555555 55000000 }
\put(823,916){\shade\ellipse{600}{300}}
\put(823,916){\ellipse{600}{300}}
\path(675,1587)(675,1062)
\path(675,1587)(675,1062)
\path(975,1587)(975,1062)
\path(975,1587)(975,1062)
\path(675,762)(675,237)
\path(675,762)(675,237)
\path(975,762)(975,237)
\path(975,762)(975,237)
\path(825,1587)(825,1062)
\path(825,1587)(825,1062)
\path(825,762)(825,237)
\path(825,762)(825,237)
\put(1500,837){\makebox(0,0){$=$}}
\put(3230,762){\makebox(0,0){$-$}}
\end{picture}
}
 \end{center}
  \caption{{\it Diagrammatic representation of $\mathcal{H}_{3,0}$ in terms of
  $\mathcal{H}_{2,0}$.}}
  \label{rel0}
\end{figure}
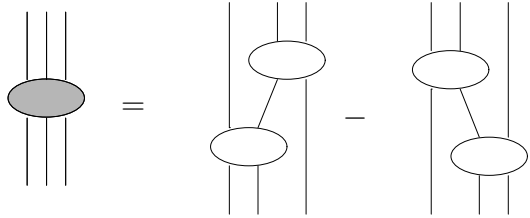

For the third charge, to zeroth order we have:
\beq
   \mathcal{H}_{3,0} = (-1)^{(C)(AB)}\word{ABC}{CAB} -
   (-1)^{(A)(BC)}\word{ABC}{BCA}
\eeq
which when written out in detail reads as:
\beqs
      \mathcal{H}_{3,0} &=& \word{abc}{cab} - \word{abc}{bca} +
      \word{ab\gamma}{\gamma ab} - \word{ab\gamma}{b\gamma a} +
      \word{a\beta c}{ca\beta} -  \word{a\beta c}{\beta ca} +\nonumber\\
     && \word{\alpha bc}{c\alpha b} -  \word{\alpha bc}{bc\alpha} -
 \word{a\beta\gamma}{\gamma a\beta} - \word{a\beta\gamma}{\beta\gamma a} -
 \word{\alpha b\gamma}{\gamma\alpha b} +\word{\alpha b\gamma}{b\gamma\alpha} +
   \nonumber\\
 &&\word{\alpha\beta c}{c\alpha\beta} +  \word{\alpha\beta c}{\beta c\alpha} +
 \word{\alpha\beta\gamma}{\gamma\alpha\beta} -
\word{\alpha\beta\gamma}{\beta\gamma\alpha}.
\eeqs
This charge is obtained by commuting two graded permutations
acting on three adjacent sites as shown in Fig.~\ref{rel0}.
As mentioned before, it is well known and easy to check that
\beq
    {[ \mathcal{H}_{2,0}, \mathcal{H}_{3,0} ]} = 0.
\eeq

We now proceed to construct the next term  $\mathcal{H}_{3,1}$
such that relation (\ref{pertcom}) is satisfied for $l=1$, namely:
\beq
    {[\mathcal{H}_{2,0}, \mathcal{H}_{3,1}]} +
    {[\mathcal{H}_{2,1}, \mathcal{H}_{3,0}]} = 0, \label{eq1}
\eeq
and such that  $\mathcal{H}_{3}(g)$ also commutes with the supercharges to
this order:
\beq
    {[{\mathcal{Q}_0}^a_\alpha, \mathcal{H}_{3,1}]} +
    {[{\mathcal{Q}_1}^a_\alpha, \mathcal{H}_{3,0}]} = 0, \label{eq2}
\eeq
This is nothing but a `brute force' computation consisting in first writing
the most general expression for the charge $\mathcal{H}_{3,1}$ and then
fixing the coefficients in such a way that (\ref{eq1}) and (\ref{eq2}) are
satisfied. The reason why we must also check (\ref{eq2}) is that in this
formulation, only the $su(2)\times su(3)$ bosonic subgroup of $su(2|3)$ is
manifestly realized while the
remaining generators will in general receive corrections, as is well known
for the case of $\mathcal{H}_2(g)$~\cite{Beisert:2003ys}.

To obtain the most general expression for $\mathcal{H}_{3,1}$ we
need to impose
the restrictions coming from \emph{manifest} $su(2)\times su(3)$ invariance and
parity. The first implies that the charge can be constructed out
of the following interactions (and their hermitian conjugates):
\beq
   \epsilon_{abc}\epsilon_{\alpha\beta}\word{Iabc}{I\alpha\beta},
   \epsilon_{abc}\epsilon_{\alpha\beta}\word{aIbc}{I\alpha\beta},
    \dots
   \epsilon_{abc}\epsilon_{\alpha\beta}\word{abcI}{\alpha\beta I},
\eeq
where the index $I$ denotes either a bosonic index or a
fermionic one and can be positioned anywhere in the upper or lower row.

However, since $\mathcal{H}_3(g)$ is parity odd,
we can only allow combinations that are odd under the parity operation
defined as:
\beq
    \mathrm{P} \word{A_1\dots A_n}{B_1\dots B_m}
    \mathrm{P}^{-1}  =   (-1)^{m+n+i(i-1)/2+j(j-1)/2}
    \word{A_n\dots A_1}{B_m\dots B_1}
\eeq
where $i$ and $j$ are respectively the number of fermionic
terms in the upper and lower rows. This restricts the form of the
charge to only ten unknown coefficients $A_1,A_2 \cdots E_1, E_2$.
The most general ansatz for $\mathcal{H}_{3,1}$ can thus be written as:
 \beqs
      \mathcal{H}_{3,1} =
    &&  A_1 \Bigg(\epsilon_{abc}\epsilon_{\alpha\beta}
                   \word{kabc}{\alpha k\beta} -
               \epsilon_{abc}\epsilon_{\alpha\beta}
                   \word{abck}{\alpha k \beta} \Bigg)+ \nonumber\\
    &&  A_2 \Bigg(\epsilon_{abc}\epsilon_{\alpha\beta}
                   \word{\gamma abc}{\alpha\gamma\beta} -
               \epsilon_{abc}\epsilon_{\alpha\beta}
                   \word{abc\gamma}{\alpha\gamma\beta}\Bigg)+\nonumber\\
    &&  B_1 \Bigg(\epsilon_{abc}\epsilon_{\alpha\beta}
                   \word{kabc}{\alpha\beta k} -
               \epsilon_{abc}\epsilon_{\alpha\beta}
                   \word{abck}{k\alpha\beta} \Bigg)+ \nonumber\\
    &&  B_2 \Bigg(\epsilon_{abc}\epsilon_{\alpha\beta}
                   \word{\gamma abc}{\alpha\beta\gamma} -
               \epsilon_{abc}\epsilon_{\alpha\beta}
                   \word{abc\gamma}{\gamma\alpha\beta}\Bigg)+\nonumber\\
    &&  C_1 \Bigg(\epsilon_{abc}\epsilon_{\alpha\beta}
                   \word{akbc}{k\alpha\beta} -
               \epsilon_{abc}\epsilon_{\alpha\beta}
                   \word{abkc}{\alpha\beta k} \Bigg)+ \nonumber\\
    &&  C_2 \Bigg(\epsilon_{abc}\epsilon_{\alpha\beta}
                   \word{a\gamma bc}{\gamma\alpha\beta} -
               \epsilon_{abc}\epsilon_{\alpha\beta}
                   \word{ab\gamma c}{\alpha\beta\gamma}\Bigg)+\nonumber\\
    &&  D_1 \Bigg(\epsilon_{abc}\epsilon_{\alpha\beta}
                   \word{akbc}{\alpha k\beta} -
               \epsilon_{abc}\epsilon_{\alpha\beta}
                   \word{abkc}{\alpha k\beta} \Bigg)+ \nonumber\\
    &&  D_2 \Bigg(\epsilon_{abc}\epsilon_{\alpha\beta}
                   \word{a\gamma bc}{\alpha\gamma\beta} -
               \epsilon_{abc}\epsilon_{\alpha\beta}
                   \word{ab\gamma c}{\alpha\gamma\beta}\Bigg)+\nonumber\\
    &&  E_1 \Bigg(\epsilon_{abc}\epsilon_{\alpha\beta}
                   \word{akbc}{\alpha\beta k} -
               \epsilon_{abc}\epsilon_{\alpha\beta}
                   \word{abkc}{k\alpha\beta} \Bigg)+ \nonumber\\
    &&  E_2 \Bigg(\epsilon_{abc}\epsilon_{\alpha\beta}
                   \word{a\gamma bc}{\alpha\beta\gamma} -
               \epsilon_{abc}\epsilon_{\alpha\beta}
                   \word{ab\gamma c}{\gamma\alpha\beta}\Bigg) +\mbox{~h.c.}.
          \label{tenterms}
\eeqs

Further possible combinations:
\beq
          \epsilon_{abc}\epsilon_{\alpha\beta}
            \word{Iabc}{I\alpha \beta} -
          \epsilon_{abc}\epsilon_{\alpha\beta}
            \word{abcI}{\alpha \beta I}
\eeq
vanish identically on graded cyclic chains.

To fix the coefficients it
is important to note that not all the ten terms in (\ref{tenterms})
are independent but that there are relations among them due to
the $su(2)\times su(3)$ invariance cutting
down on the number of independent coefficients. The
first relation that one can utilize is that, for any $I$, $J$ and $K$
\beq
\epsilon_{abc}\Bigg(
                   \word{kabc}{IJK} - \word{abck}{IJK}  -
                   \word{akbc}{IJK} + \word{abkc}{IJK} \Bigg) = 0.
\eeq This relation is specific to $su(3)$ and can be proved by
looking at the action of the l.h.s. on any single trace state.
On a given single trace state, the l.h.s. will give a non-zero
contribution only on parts of the state that contain four
consecutive bosonic spins. Furthermore,  because of the epsilon
tensor, three of those spins have to be different. It is now
straightforward to convince oneself that on any such 'block' of
four bosonic spins the l.h.s. always sums up to zero. This
relation immediately implies that $D_1$ and $B_1$ can be
eliminated or, in other words, only the combinations $A_1 + D_1$,
$B_1 + C_1$ and $B_1 + E_1$ appear in the ansatz for
$\mathcal{H}_{3,1}$. Acting on bosonic states with (\ref{eq1})
then immediately fixes $B_1+E_1 =0$, $A_1+D_1 =-1$ and $B_1+C_1
=1$.

There is a similar relation, this time specific to $su(2)$, that
allows one to eliminate some of the coefficients in the fermionic
sector. Namely, for any $I$, $J$, $K$ and $L$ \beq
    \epsilon_{\alpha\beta}
                 \Bigg( \word{IJKL}{\gamma\alpha\beta}+
                   \word{IJKL}{\alpha\beta\gamma} -
                   \word{IJKL}{\alpha\gamma\beta} \Bigg) = 0.
\eeq
This relation implies that
the term multiplying $D_2$ can be eliminated and the coefficients
$C_2$ and $E_2$ change to $C_2 +D_2$ and $E_2+D_2$ respectively.

After that, requiring (\ref{eq1}) to be satisfied on fermionic states,
we obtain
$A_2=1$, $B_2 =0$, $C_2+D_2 =1$ and $E_2+D_2 =0$, which fixes \emph{all} the
coefficients.
The final formula for $\mathcal{H}_{3,1}$ thus takes
the form:
\beqs
         \mathcal{H}_{3,1}=\!\!\!\!\!\!&&\epsilon_{abc}\epsilon_{\alpha\beta}
                   \Bigg(
                   - \word{kabc}{\alpha k\beta}
                   + \word{abck}{\alpha k \beta}
                   + \word{akbc}{k\alpha\beta}
                   - \word{abkc}{\alpha\beta k} \Bigg) +\nonumber\\
                   && \epsilon_{abc}\epsilon_{\alpha\beta}\Bigg(
                   + \word{\gamma abc}{\alpha\gamma\beta}
                   - \word{abc\gamma}{\alpha\gamma\beta}
                   + \word{a\gamma bc}{\gamma\alpha\beta}
                   - \word{ab\gamma c}{\alpha\beta\gamma}\Bigg) +
                   \mbox{h.c.} \label{final}
\eeqs
It can also be checked that (\ref{eq2}) does not introduce any
further constraint, i.e. supersymmetry is preserved as well.
\section{Concluding Remarks}
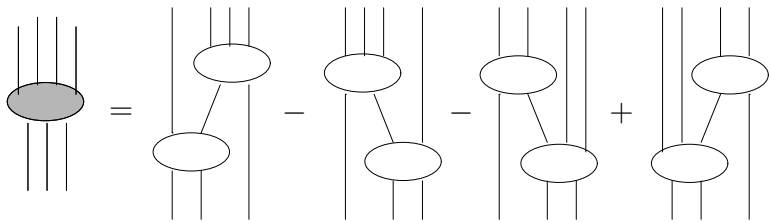
\begin{figure}[hbp]
  \begin{center}
\setlength{\unitlength}{0.00066667in}
\begingroup\makeatletter\ifx\SetFigFont\undefined%
\gdef\SetFigFont#1#2#3#4#5{%
  \reset@font\fontsize{#1}{#2pt}%
  \fontfamily{#3}\fontseries{#4}\fontshape{#5}%
  \selectfont}%
\fi\endgroup%
{\renewcommand{\dashlinestretch}{30}
\begin{picture}(5955,1689)(0,-10)
\texture{88555555 55000000 555555 55000000 555555 55000000 555555 55000000
    555555 55000000 555555 55000000 555555 55000000 555555 55000000
    555555 55000000 555555 55000000 555555 55000000 555555 55000000
    555555 55000000 555555 55000000 555555 55000000 555555 55000000 }
\put(308,928){\shade\ellipse{600}{300}}
\put(308,928){\ellipse{600}{300}}
\path(172,762)(172,237)
\path(172,762)(172,237)
\path(472,762)(472,237)
\path(472,762)(472,237)
\path(322,762)(322,237)
\path(322,762)(322,237)
\path(247,1587)(247,1062)
\path(247,1587)(247,1062)
\path(547,1512)(547,987)
\path(547,1512)(547,987)
\path(397,1587)(397,1062)
\path(397,1587)(397,1062)
\path(97,1512)(97,987)
\path(97,1512)(97,987)
\put(1447,537){\ellipse{600}{300}}
\put(1763,1228){\ellipse{600}{300}}
\path(1297,387)(1297,12)
\drawline(1297,687)(1297,687)
\path(1297,687)(1297,1662)
\path(1522,387)(1522,12)
\path(1672,1062)(1522,687)
\path(1897,1062)(1897,12)
\path(1897,1362)(1897,1662)
\path(1597,1362)(1597,1662)
\path(1747,1362)(1747,1662)
\put(3097,462){\ellipse{600}{300}}
\put(2781,1153){\ellipse{600}{300}}
\drawline(2647,987)(2647,987)
\path(2647,987)(2647,12)
\path(3022,612)(2872,987)
\path(3022,312)(3022,12)
\path(3247,612)(3247,1662)
\path(3247,312)(3247,12)
\path(2947,1287)(2947,1662)
\path(2797,1287)(2797,1662)
\path(2647,1287)(2647,1662)
\put(3997,1137){\ellipse{600}{300}}
\put(4313,446){\ellipse{600}{300}}
\path(3847,1287)(3847,1662)
\drawline(3847,987)(3847,987)
\path(3847,987)(3847,12)
\path(4072,1287)(4072,1662)
\path(4222,612)(4072,987)
\path(4447,312)(4447,12)
\path(4522,537)(4522,1662)
\path(4372,612)(4372,1662)
\path(4222,312)(4222,12)
\put(5647,1137){\ellipse{600}{300}}
\put(5331,446){\ellipse{600}{300}}
\path(5797,1287)(5797,1662)
\drawline(5797,987)(5797,987)
\path(5797,987)(5797,12)
\path(5572,1287)(5572,1662)
\path(5422,612)(5572,987)
\path(5197,312)(5197,12)
\path(5122,537)(5122,1662)
\path(5272,612)(5272,1662)
\path(5422,312)(5422,12)
\put(900,837){\makebox(0,0){$=$}}
\put(2250,837){\makebox(0,0){$-$}}
\put(3550,837){\makebox(0,0){$-$}}
\put(4800,837){\makebox(0,0){$+$}}
\end{picture}
}
 \end{center}
  \caption{{\it Diagrammatic representation of $\mathcal{H}_{3,1}$ in terms of
  $\mathcal{H}_{2,0}$ and  $\mathcal{H}_{2,1}$. Each term on the r.h.s. 
corresponds to a pair of terms (bosonic and fermionic) in eq.~(\ref{final}).
In principle, there could be two more diagrams
like the last two but with $\mathcal{H}_{2,0}$ and $\mathcal{H}_{2,1}$ 
sharing two legs instead of one. These two diagrams cancel out in this case.
}}
  \label{rel1}
\end{figure}

Eq. (\ref{final}) is suggestive because it shows that $\mathcal{H}_{3,1}$
can be expressed in terms of the interactions already present in the
Hamiltonian by a rather simple generalization of Fig.~\ref{rel0} shown
diagrammatically in Fig.~\ref{rel1}.

Having a diagrammatic approach at hand, it is conceivable
that one may be able to utilize the intuition generated by the
present work and construct the complete monodromy matrix, which
generates all the conserved quantities for this spin chain. Such a
construction could then be used to derive the Bethe equations for
the dynamical spin chain and establish  complete integrability for
this many body problem.
By some preliminary analysis, we are quite convinced that
such simple structure generalizes to higher loops and higher
charges~\footnote{After the first version of this paper was submitted to the
{\tt arXive} we have completed the construction of $\mathcal{H}_{3,2}$, see
Appendix.}. This is yet another encouraging sign that it may be
possible to find a transfer matrix even for dynamical spin chains
of this type.

One further interesting fact is that diagrammatic relations such
as Fig.~\ref{rel0} and Fig.~\ref{rel1} do not know about the
details of the underlying model and can in principle be applied to
the full $psu(2,2|4)$ chain perhaps even allowing one to fix the
coefficients of the dilatation operator to higher loops.

Apart from local conserved charges, complete integrability
would also require an understanding of how the Yangian symmetry of
the gauge theory dilatation operator may be realized in the
context of the dynamical spin chains. This symmetry is known to be
present, at least to the first few orders in perturbation theory,
in many of the sub-sectors of the gauge theory that do not exhibit
the dynamical behavior
\cite{Vallilo:2003nx,Dolan:2003uh,Alday:2003zb,
Agarwal:2004sz, Wolf:2004hp, Agarwal:2005ed}, 
and it is also known to be present
in the full gauge theory dilatation operator at one loop~\cite{Dolan:2004ys}. 
Hence, a
novel realization of Yangian symmetry, relevant for dynamical spin
chains, remains a crucial piece of the puzzle to be discovered.

Finally, even for theories with less on no supersymmetries, it may
be interesting to try deforming the chiral sector discussed in
\cite{Ferretti:2004ba,Beisert:2004fv} in a way that preserves
integrability. This sector is known not to be closed beyond one
loop but, given the fact that the antichiral `impurities' appear
with a mass gap in long operators it might be possible to develop
a truncation scheme in which integrability generalizes to higher
loops even in these cases.

\section*{Acknowledgments}

We would like to thank R.~Argurio, M.~Bertolini, R.~Heise, T.~Klose,
J.~Minahan, S.~Rajeev and K.~Zarembo for discussions. Special thanks go to
N.~Beisert for many discussions and for
not believing us when we were getting the wrong results!
G.F. would like to thank the University of Rochester for its warm hospitality
while this work was initiated.
The research of G.F. is supported by the Swedish Research Council
(Vetenskapsr{\aa}det) contracts 622-2003-1124 and 621-2002-3884.
\vfill\eject

\section*{Appendix}

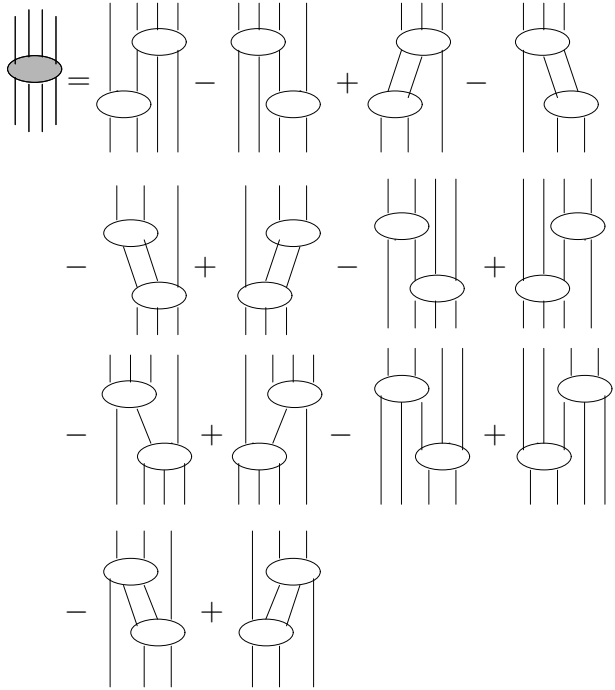
\begin{figure}[hbp]
  \begin{center}
\setlength{\unitlength}{0.00046667in}
\begingroup\makeatletter\ifx\SetFigFont\undefined%
\gdef\SetFigFont#1#2#3#4#5{%
  \reset@font\fontsize{#1}{#2pt}%
  \fontfamily{#3}\fontseries{#4}\fontshape{#5}%
  \selectfont}%
\fi\endgroup%
{\renewcommand{\dashlinestretch}{30}
\begin{picture}(6705,7614)(0,-10)
\put(1688,7153){\ellipse{600}{300}}
\put(1297,6462){\ellipse{600}{300}}
\drawline(1222,6612)(1222,6612)
\path(1672,6987)(1672,5937)
\path(1897,7062)(1897,6012)(1897,5937)
\path(1672,7287)(1672,7587)
\path(1447,7287)(1447,7587)
\path(1897,7287)(1897,7587)
\path(1447,7062)(1447,6612)
\path(1147,6312)(1147,5937)
\path(1147,6612)(1147,7587)
\path(1447,6312)(1447,5937)
\put(2781,7153){\ellipse{600}{300}}
\put(3172,6462){\ellipse{600}{300}}
\drawline(3247,6612)(3247,6612)
\path(2797,6987)(2797,5937)
\path(2572,7062)(2572,6012)(2572,5937)
\path(2797,7287)(2797,7587)
\path(3022,7287)(3022,7587)
\path(2572,7287)(2572,7587)
\path(3022,7062)(3022,6612)
\path(3322,6312)(3322,5937)
\path(3322,6612)(3322,7587)
\path(3022,6312)(3022,5937)
\put(4613,7153){\ellipse{600}{300}}
\put(4297,6462){\ellipse{600}{300}}
\path(4822,7062)(4822,6012)(4822,5937)
\path(4597,7287)(4597,7587)
\path(4372,7287)(4372,7587)
\path(4822,7287)(4822,7587)
\path(4597,6987)(4447,6537)
\path(4372,7062)(4222,6612)
\path(4447,6312)(4447,5937)
\path(4147,6312)(4147,5937)
\put(5931,7153){\ellipse{600}{300}}
\put(6247,6462){\ellipse{600}{300}}
\path(5722,7062)(5722,6012)(5722,5937)
\path(5947,7287)(5947,7587)
\path(6172,7287)(6172,7587)
\path(5722,7287)(5722,7587)
\path(5947,6987)(6097,6537)
\path(6172,7062)(6322,6612)
\path(6097,6312)(6097,5937)
\path(6397,6312)(6397,5937)
\put(1688,4346){\ellipse{600}{300}}
\put(1372,5037){\ellipse{600}{300}}
\path(1897,4437)(1897,5487)(1897,5562)
\path(1672,4212)(1672,3912)
\path(1447,4212)(1447,3912)
\path(1897,4212)(1897,3912)
\path(1672,4512)(1522,4962)
\path(1447,4437)(1297,4887)
\path(1522,5187)(1522,5562)
\path(1222,5187)(1222,5562)
\put(2856,4346){\ellipse{600}{300}}
\put(3172,5037){\ellipse{600}{300}}
\path(2647,4437)(2647,5487)(2647,5562)
\path(2872,4212)(2872,3912)
\path(3097,4212)(3097,3912)
\path(2647,4212)(2647,3912)
\path(2872,4512)(3022,4962)
\path(3097,4437)(3247,4887)
\path(3022,5187)(3022,5562)
\path(3322,5187)(3322,5562)
\put(4763,4421){\ellipse{600}{300}}
\put(4372,5112){\ellipse{600}{300}}
\drawline(4297,4962)(4297,4962)
\path(4747,4587)(4747,5637)
\path(4972,4512)(4972,5562)(4972,5637)
\path(4747,4287)(4747,3987)
\path(4522,4287)(4522,3987)
\path(4972,4287)(4972,3987)
\path(4522,4512)(4522,4962)
\path(4222,5262)(4222,5637)
\path(4222,4962)(4222,3987)
\path(4522,5262)(4522,5637)
\put(5931,4421){\ellipse{600}{300}}
\put(6322,5112){\ellipse{600}{300}}
\drawline(6397,4962)(6397,4962)
\path(5947,4587)(5947,5637)
\path(5722,4512)(5722,5562)(5722,5637)
\path(5947,4287)(5947,3987)
\path(6172,4287)(6172,3987)
\path(5722,4287)(5722,3987)
\path(6172,4512)(6172,4962)
\path(6472,5262)(6472,5637)
\path(6472,4962)(6472,3987)
\path(6172,5262)(6172,5637)
\put(1356,3253){\ellipse{600}{300}}
\put(1747,2562){\ellipse{600}{300}}
\drawline(1822,2712)(1822,2712)
\path(1372,3387)(1372,3687)
\path(1597,3387)(1597,3687)
\path(1147,3387)(1147,3687)
\path(1897,2712)(1897,3687)
\path(1447,3087)(1597,2712)
\path(1222,3087)(1222,2037)
\path(1522,2487)(1522,2037)
\path(1972,2487)(1972,2037)
\path(1747,2412)(1747,2037)
\put(3188,3253){\ellipse{600}{300}}
\put(2797,2562){\ellipse{600}{300}}
\drawline(2722,2712)(2722,2712)
\path(3172,3387)(3172,3687)
\path(2947,3387)(2947,3687)
\path(3397,3387)(3397,3687)
\path(2647,2712)(2647,3687)
\path(3097,3087)(2947,2712)
\path(3322,3087)(3322,2037)
\path(3022,2487)(3022,2037)
\path(2572,2487)(2572,2037)
\path(2797,2412)(2797,2037)
\put(4822,2562){\ellipse{600}{300}}
\put(4372,3312){\ellipse{600}{300}}
\drawline(4897,2712)(4897,2712)
\path(4972,2412)(4972,2037)
\path(4672,2412)(4672,2037)
\path(4597,3162)(4597,2637)
\path(4822,2712)(4822,3762)
\path(5047,2637)(5047,3762)
\path(4372,3162)(4372,2037)
\path(4222,3462)(4222,3762)
\path(4522,3462)(4522,3762)
\path(4147,3237)(4147,2037)
\put(5947,2562){\ellipse{600}{300}}
\put(6397,3312){\ellipse{600}{300}}
\drawline(5872,2712)(5872,2712)
\path(5797,2412)(5797,2037)
\path(6097,2412)(6097,2037)
\path(6172,3162)(6172,2637)
\path(5947,2712)(5947,3762)
\path(5722,2637)(5722,3762)
\path(6397,3162)(6397,2037)
\path(6547,3462)(6547,3762)
\path(6247,3462)(6247,3762)
\path(6622,3237)(6622,2037)
\texture{88555555 55000000 555555 55000000 555555 55000000 555555 55000000 
	555555 55000000 555555 55000000 555555 55000000 555555 55000000 
	555555 55000000 555555 55000000 555555 55000000 555555 55000000 
	555555 55000000 555555 55000000 555555 55000000 555555 55000000 }
\put(308,6853){\shade\ellipse{600}{300}}
\put(308,6853){\ellipse{600}{300}}
\put(1372,1287){\ellipse{600}{300}}
\put(1672,612){\ellipse{600}{300}}
\put(3172,1287){\ellipse{600}{300}}
\put(2872,612){\ellipse{600}{300}}
\path(247,7512)(247,6987)
\path(247,7512)(247,6987)
\path(547,7437)(547,6912)
\path(547,7437)(547,6912)
\path(397,7512)(397,6987)
\path(397,7512)(397,6987)
\path(97,7437)(97,6912)
\path(97,7437)(97,6912)
\path(97,6762)(97,6237)
\path(97,6762)(97,6237)
\path(547,6762)(547,6237)
\path(547,6762)(547,6237)
\path(397,6687)(397,6162)
\path(397,6687)(397,6162)
\path(247,6687)(247,6162)
\path(247,6687)(247,6162)
\path(1222,1437)(1222,1737)
\path(1522,1437)(1522,1737)
\path(1147,1212)(1147,12)
\path(1297,1137)(1447,687)
\path(1522,1137)(1672,762)
\path(1822,762)(1822,1737)
\path(1522,462)(1522,12)
\path(1822,462)(1822,12)
\path(3322,1437)(3322,1737)
\path(3022,1437)(3022,1737)
\path(3397,1212)(3397,12)
\path(3247,1137)(3097,687)
\path(3022,1137)(2872,762)
\path(2722,762)(2722,1737)
\path(3022,462)(3022,12)
\path(2722,462)(2722,12)
\put(800,6700){\makebox(0,0){$=$}}
\put(2197,6700){\makebox(0,0){$-$}}
\put(3772,6700){\makebox(0,0){$+$}}
\put(5200,6700){\makebox(0,0){$-$}}
\put(772,4662){\makebox(0,0){$-$}}
\put(2197,4662){\makebox(0,0){$+$}}
\put(3772,4662){\makebox(0,0){$-$}}
\put(5400,4662){\makebox(0,0){$+$}}
\put(772,2800){\makebox(0,0){$-$}}
\put(2272,2800){\makebox(0,0){$+$}}
\put(3697,2800){\makebox(0,0){$-$}}
\put(5400,2800){\makebox(0,0){$+$}}
\put(772,837){\makebox(0,0){$-$}}
\put(2272,837){\makebox(0,0){$+$}}
\end{picture}
}
 \end{center}
  \caption{{\it Diagrammatic representation of $\mathcal{H}_{3,2}$ in terms of
  $\mathcal{H}_{2,0}$, $\mathcal{H}_{2,1}$ and $\mathcal{H}_{2,2}$. 
  Diagrams with less incoming and outgoing legs should be interpreted as
  having an extra spectator leg.
}}
  \label{rel2}
\end{figure}

In this Appendix we simply present the diagrammatic expression 
(Fig.~\ref{rel2}) of 
$\mathcal{H}_{3,2}$ in terms of $\mathcal{H}_{2,0}$,  
$\mathcal{H}_{2,1}$ and $\mathcal{H}_{2,2}$. In general, it should be 
possible to obtain the $k$-th loop contribution $\mathcal{H}_{3,k}$ in
terms of bilinears $\mathcal{H}_{2,i} \mathcal{H}_{2,k-i}$ 
for $i=0\dots k$.

Imposing that the integrability condition is obeyed by 
$\mathcal{H}_{3,2}$ fixes the values of the undetermined parameter 
$\delta_2$ in~\cite{Beisert:2003ys}. ($\delta_1$ is still undetermined 
and $\delta_3$ is not a true free 
parameter since it multiplies a quantity that is identically zero.) 
With the normalizations as 
in~\cite{Beisert:2003ys}, ($\alpha_1 = 1$ and $\alpha_3 = 0$),
setting the rotation angles $\gamma_i = 0$ we find
$\delta_2 = -5/4$.
\vfill\eject


\begin{thebibliography}{99}

\bibitem{Minahan:2002ve}
  J.~A.~Minahan and K.~Zarembo,
  JHEP {\bf 0303}, 013 (2003)
  [arXiv:hep-th/0212208].

\bibitem{Beisert:2003yb}
  N.~Beisert and M.~Staudacher,
  Nucl.\ Phys.\ B {\bf 670}, 439 (2003)
  [arXiv:hep-th/0307042].

\bibitem{Bukhvostov:1985rn}
  A.~P.~Bukhvostov, G.~V.~Frolov, L.~N.~Lipatov and E.~A.~Kuraev,
  Nucl.\ Phys.\ B {\bf 258}, 601 (1985).

\bibitem{Lipatov:1993yb}
  L.~N.~Lipatov,
  arXiv:hep-th/9311037.

\bibitem{Faddeev:1994zg}
  L.~D.~Faddeev and G.~P.~Korchemsky,
  Phys.\ Lett.\ B {\bf 342}, 311 (1995)
  [arXiv:hep-th/9404173].

\bibitem{Braun:2003rp}
  V.~M.~Braun, G.~P.~Korchemsky and D.~Muller,
  Prog.\ Part.\ Nucl.\ Phys.\  {\bf 51}, 311 (2003)
  [arXiv:hep-ph/0306057].

\bibitem{Ferretti:2004ba}
  G.~Ferretti, R.~Heise and K.~Zarembo,
  Phys.\ Rev.\ D {\bf 70}, 074024 (2004)
  [arXiv:hep-th/0404187].

\bibitem{Belitsky:2004cz}
  A.~V.~Belitsky, V.~M.~Braun, A.~S.~Gorsky and G.~P.~Korchemsky,
  Int.\ J.\ Mod.\ Phys.\ A {\bf 19}, 4715 (2004)
  [arXiv:hep-th/0407232].

\bibitem{Belitsky:2004sc}
  A.~V.~Belitsky, S.~E.~Derkachov, G.~P.~Korchemsky and A.~N.~Manashov,
  Nucl.\ Phys.\ B {\bf 708}, 115 (2005)
  [arXiv:hep-th/0409120].

\bibitem{Beisert:2004fv}
  N.~Beisert, G.~Ferretti, R.~Heise and K.~Zarembo,
  Nucl.\ Phys.\ B {\bf 717}, 137 (2005)
  [arXiv:hep-th/0412029].

\bibitem{Beisert:2003ys}
  N.~Beisert,
  Nucl.\ Phys.\ B {\bf 682}, 487 (2004)
  [arXiv:hep-th/0310252].

\bibitem{Serban:2004jf}
  D.~Serban and M.~Staudacher,
  JHEP {\bf 0406}, 001 (2004)
  [arXiv:hep-th/0401057].

\bibitem{Beisert:2004hm}
  N.~Beisert, V.~Dippel and M.~Staudacher,
  JHEP {\bf 0407}, 075 (2004)
  [arXiv:hep-th/0405001].

\bibitem{Beisert:2005fw}
  N.~Beisert and M.~Staudacher,
  arXiv:hep-th/0504190.

\bibitem{Staudacher:2004tk}
  M.~Staudacher,
  JHEP {\bf 0505}, 054 (2005)
  [arXiv:hep-th/0412188].

\bibitem{Haldane:1992sj}
  F.~D.~M.~Haldane, Z.~N.~C.~Ha, J.~C.~Talstra, D.~Bernard and V.~Pasquier,
  Phys.\ Rev.\ Lett.\  {\bf 69}, 2021 (1992).

\bibitem{Bernard:1993uz}
  D.~Bernard, M.~Gaudin, F.~D.~M.~Haldane and V.~Pasquier,
  arXiv:hep-th/9301084.

\bibitem{Bernard:1993va}
  D.~Bernard, M.~Gaudin, F.~D.~M.~Haldane and V.~Pasquier,
  J.\ Phys.\ A {\bf 26}, 5219 (1993).

\bibitem{Inozemtsev:2002vb}
  V.~I.~Inozemtsev,
  Phys.\ Part.\ Nucl.\  {\bf 34}, 166 (2003)
  [Fiz.\ Elem.\ Chast.\ Atom.\ Yadra {\bf 34}, 332 (2003)]
  [arXiv:hep-th/0201001].

\bibitem{Grabowski:1994rb}
  M.~P.~Grabowski and P.~Mathieu,
  J.\ Phys.\ A {\bf 28}, 4777 (1995)
  [arXiv:hep-th/9412039].

\bibitem{Arutyunov:2003rg}
  G.~Arutyunov and M.~Staudacher,
  JHEP {\bf 0403}, 004 (2004)
  [arXiv:hep-th/0310182].

\bibitem{Engquist:2004bx}
  J.~Engquist,
  JHEP {\bf 0404}, 002 (2004)
  [arXiv:hep-th/0402092].

\bibitem{Beisert:2003tq}
  N.~Beisert, C.~Kristjansen and M.~Staudacher,
  Nucl.\ Phys.\ B {\bf 664}, 131 (2003)
  [arXiv:hep-th/0303060].

\bibitem{Essler:1992uc}
  F.~H.~L.~Essler, V.~E.~Korepin and K.~Schoutens,\hfill\break
  arXiv:cond-mat/9211001.

\bibitem{Vallilo:2003nx}
  B.~C.~Vallilo,
  JHEP {\bf 0403}, 037 (2004)
  [arXiv:hep-th/0307018].

\bibitem{Dolan:2003uh}
  L.~Dolan, C.~R.~Nappi and E.~Witten,
  JHEP {\bf 0310}, 017 (2003)
  [arXiv:hep-th/0308089].

\bibitem{Alday:2003zb}
  L.~F.~Alday,
  JHEP {\bf 0312}, 033 (2003)
  [arXiv:hep-th/0310146].

\bibitem{Agarwal:2004sz}
  A.~Agarwal and S.~G.~Rajeev,
  arXiv:hep-th/0409180.

\bibitem{Wolf:2004hp}
  M.~Wolf,
  JHEP {\bf 0502}, 018 (2005)
  [arXiv:hep-th/0412163].

\bibitem{Agarwal:2005ed}
  A.~Agarwal,
  arXiv:hep-th/0506095.

\bibitem{Dolan:2004ys}
  L.~Dolan and C.~R.~Nappi,
  Nucl.\ Phys.\ B {\bf 717}, 361 (2005)
  [arXiv:hep-th/0411020].

\end{thebibliography}
\end{document}